\documentclass[final,3p,times,twocolumn]{elsarticle}
\usepackage{booktabs}
\usepackage[font=small,labelfont=bf]{caption}

\usepackage{graphicx}
\usepackage{color,soul}
\usepackage{amssymb}
\usepackage{lineno}

\journal{Biosystems}
\begin{document}
\begin{frontmatter}
\title{The distribution of information for sEMG signals in the rectal cancer treatment process}

\author[us]{Paulina Trybek}
\author[uj]{Michal Nowakowski}
\author[sk]{Jerzy Salowka}
\author[us]{Lukasz Machura\corref{cor1}}
\ead{lukasz.machura@smcebi.edu.pl}
\cortext[cor1]{tel +48 32 349 76 05}

\address[us]{Division of Computational Physics and Electronics, Institute of Physics, 
Silesian Centre for Education and Interdisciplinary Research, University of Silesia in Katowice, Poland}
\address[uj]{Department of Medical Education, Jagiellonian University Medical College, Krakow, Poland}
\address[sk]{Department of Surgery, Stanley Dudrick Memorial Hospital, Skawina, Poland}

\begin{abstract}
The electrical activity of external anal sphincter can be registered with surface electromyography. This signals are known to be highly complex and nonlinear. This work aims in characterisation of the information carried in the signals by harvesting the concept of information entropy. We will focus of two classical measures of the complexity. Firstly the Shannon entropy is addressed. It is related to the probability spectrum of the possible states. Secondly the Spectral entropy is described, as a simple frequency-domain analog of the time-domain Shannon characteristics. We discuss the power spectra for separate time scales and present the characteristics which can represent the dynamics of electrical activity of this specific muscle group. 
We find that the rest and maximum contraction states represent rather different spectral characteristic of entropy, with close-to-normal contraction and negatively skewed rest state.
\end{abstract}


\end{frontmatter}

%\linenumbers

\section{Introduction}
\label{intro}
The colorectal cancer (CRC) remains one of the most common cancer with a high mortality rate \cite{ferlay2015cancer, arnold2016global}. The appropriate diagnostic procedures are required for both prior to treatment and during the multimodal therapy. 
In order to implement the most effective healing process while increasing the probability of fast recovery, not only the tumor diagnostic techniques are important but also the monitoring of surrounding anatomical structures that may be potentially affected by the invasive therapy including surgery or radiation \cite{ammann2003impact, ridolfi2016low}. 
On top of that the understanding of the evolution and mutation of the CRC cells are crucial for more effective future treatment \cite{roerink2018intra}.
In this work we study the results of some less conventional method of diagnosis of an external anal sphincter (EAS) activity before and after the surgery. 
{The classification of patients can be effectively based on the anorectal manometry together with the descriptive parameters like root mean square or median frequency of raw sEMG signals} \cite{nowakowski2017sensitivity}.
The electrical activity of this muscle group is investigated through the surface electromyography technique (sEMG) before and after the anterior resection (AR) or the lower anterior resection (LAR).
The registered time series indicate a high degree of complexity, hence it is necessary to apply methods that adequately characterise the dynamics of the complex process hidden in the high-dimensional data sets. 
{The sEMG signals in question exhibit two distinct scaling regions which can be identified with multifractal spectra. 
The multi-- and mono--fractal nature of the signals can be seen for the short and large time scales respectively with the former being the result of the long--range correlations for weak and large fluctuations} \cite{trybek2018multifractal}.
Quite recently, considerable attention has been paid to the entropy-based techniques with a very promising potential in bio-medical signal processing \cite{nikulin2004comment,costa2005multiscale,kaplanis2010multiscale,borowska2015entropy}. 
The idea of implementing the entropy notions for the measure of biodiversity is present in an increasing number of the studied biosystems, from the microscopic level \cite{stewart1997shannon, zhang2009empirical} to the whole body response that can be investigated by the spectrum of electrophysiological signals as electrocardiography (ECG) \cite{cysarz2000entropies,kamath2012entropy,makowiec2015entropic}, encephalography (EEG) \cite{phung2014using} or magnetoencephalography (MEG) \cite{gomez2010entropy}. 
{The specific values of the approximate entropy measures and their dependence on the time scales were analysed for the sEMG signals registered with patients with colorectal cancer} \cite{trybek2018sample}. 
{The statistically significant differences among all stages of medical treatment and for all consecutive depths of rectum area were found for the Sample Entropy ($SampEn$)} \cite{richman2000physiological}. 
{The analysis of $SampEn$ over multiple time scales exhibits the most visible differences between AR and LAR groups were identified one month after an operation. 
It was also shown that the information carried out by the sEMG signals measured one year after the surgery returns to the state of that before the surgery for the selected cases} \cite{trybek2018sample}.

This work presents the application of basics ideas taken from the information theory for the characterisation of a specific neuromuscular activity of EAS. 
To better understand of features of analysed signals and how complexity is assessed from them the standard Shannon entropy ($H_x$) with their frequency analogue -- spectral entropy ($H_f$) were applied. 
A thorough spectral analysis was also carried out for the full characterisation of the examined time series.  
 
\section{Material and methods}\label{mm} 
\textbf{Patients.} The study group include 20 subjects. Among them were 6 female and 14 male with the average age $64.6$. All patients were diagnosed with a rectal cancer and undergone one of a surgical procedure (LAR or AR). 

\textbf{Data acquisition.} The series were recorded before surgery, and two times after, respectively 6 months and 1 year in the postoperative period. In addition the two extreme cases of muscle tension, relaxation and maximum voluntary contraction (MVC) were considered separately.
The measuring device developed at the Politechnico di Torino in collaboration with the OT-Bioelettronica consist of 3 rings of 16 silver/silver oxide bar electrodes.  The proper measurement took 10 seconds each. 
The sampling frequency was set to 2048 Hz. 
Low and high pass filters were used at 10 and 500 Hz respectively and resulted in typical 3dB bandwidth for the Analog to Digital Converter.
{The 20 subjects x 16 channels x 3 different depth of rectum x 2 stages of muscle tension (REST vs MVC) x 3 different stages of treatment resulted in about 6000 signals for analysis.}

\begin{figure}
    \centering
    \includegraphics[width=0.8\linewidth]{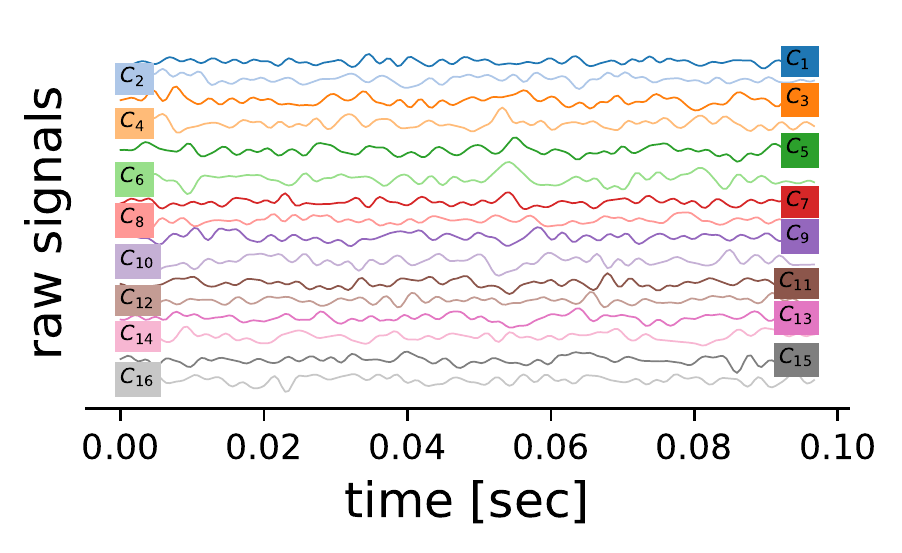}
    \caption{The raw signals registered at 16 electrodes at anal channel depth of 5cm was truncated to 100 milliseconds for visibility.}
    \label{data}
\end{figure}

\section{Information entropy}
In statistical thermodynamics the macroscopic state of the microsystem is characterised by the distribution of the possible states. 
The entropy of distribution $p_i$ of a discrete set $\{X_i\}_{i=1}^n$ of $n$ microstates is given by the Gibbs formula $S = -k_B T \sum_i p(X_i) \log p(X_i)$, where $T$ is a temperature and $k_B$ is the Boltzmann constant. 
This definition works also for systems far away from equilibrium.
An early idea of Gibbs was transmitted to the information theory in 1948 by Claude Shannon \cite{shannon1948mathematical} and since then serves as a popular definition of the complexity of signals \cite{mackay2003information}.

\subsection{Shannon Entropy}
{The average Shannon information content $H_x$ of an outcome $x$} characterises the average information contained in the signal. 
It calculates the number of possible states that the system represented by a respective time series can take. 
$H_x$ is often associated with the degree of uncertainty present in the signal or with the information that is inversely proportional to its predictability. 
The basic formula for calculating $H_x$ of time series of $N$ samples $\{x_i\}_{i=1}^N$ with the probability distribution function $p(x_i)$ is given as a average number of the logarithm of the probability of a certain state $x_i$.
\begin{equation}
 H_x = -\sum_{i=1}^{N} p(x_i) \log p(x_i)
 \label{shannon_eq}
\end{equation}
The higher values of $H_x$ are determined by more unpredictable character of the series. 
The maximum of $H_x$ occurs in the case when all the states are equally probable.
The base for logarithm in (\ref{shannon_eq}) defines the unit of entropy. Throughout this paper we will use the original formulation of Shannon \cite{shannon1948mathematical}, where base equals 2 and the unit of entropy is called shannon $Sh$ or bit.
\begin{figure}[htbp]
    \centering
    \includegraphics[width=1.\linewidth]{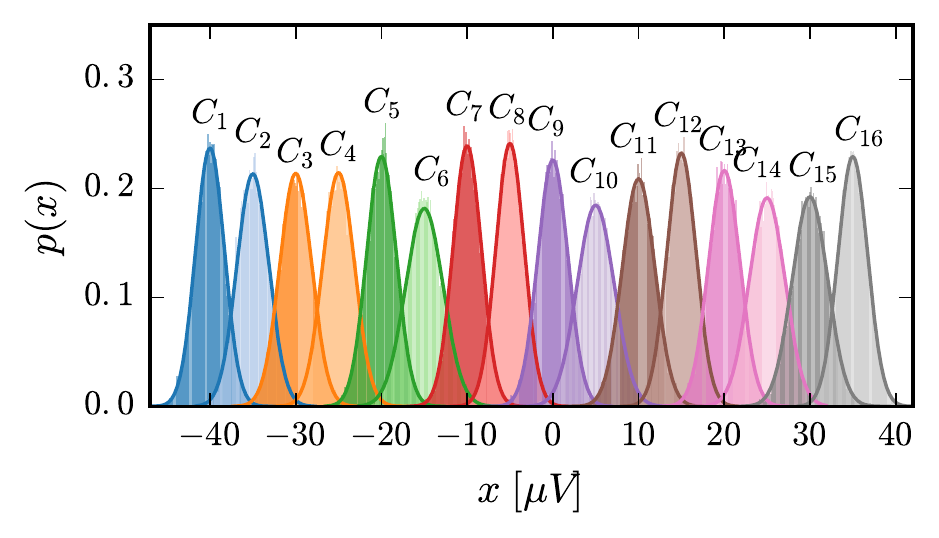}
    \caption{The probability distributions of amplitudes for 16 electrodes, shifted along abscissa for visibility. State $x$ represent the voltage registered with sEMG.}
    \label{hist}
\end{figure}

\subsection{Power Spectral Entropy}
{The Power Spectral Entropy} ($H_f$) is a simple analog of the $H_x$ in the frequency--domain. 
It tells us how broadly the power in the signal is distributed across the available frequencies. It is used as a measure of the width and uniformity of the power spectrum. 
The frequency components of the power spectrum are always found to be linearly independent, regardless of the presence or absence of correlations. 
It gives an advantage over the standard method of creating a histogram where the correlations present in the signal are lost.
The main limitation of this method lies in the fact that the determination of the power spectrum filters out all the nonlinearities that might be present in the signal, thus is an
insensitive method for more subtle features of the signals \cite{semmlow2014biosignal}. 
The first step in the procedure of estimation of the $H_f$ is the calculation of the Power Spectral Density $S_{xx}$ function. Next, the $S_{xx}$ is normalised by the sum of all its spectral components $s_i$.
\begin{equation}
 S_{xx}(f_i) = \frac{1}{N}|x(f_i)|^2,  
 \qquad
 s_i = \frac{S_{xx}(f_i)}{\sum_i S_{xx}(f_i)}  
 \label{psd}
\end{equation}
{Above $x(f_i)$ is the Fourier transform (FT) of the process $x(t)$ and $f$ is the frequency in $Hz$. 
For the estimation of the PSD of $x(t)$ we use the periodogram} \cite{semmlow2014biosignal}. {As a non-parametric estimator of the PSD is not always consistent due to the fact that its variance does not always converge to zero with the signal length, for formal analysis of the PSD (see Fig,} \ref{periodograms}) {we use the Welch method} \cite{welch1967use}. {In short, it splits the random signal into overlapping segments, estimate the PSD for each one and returns the average over these local estimates.}
The Power Spectral Entropy is defined as
\begin{equation}
 H_f =  -\sum_i s_i \log s_i 
 \label{pse}
\end{equation}
If only few frequencies are present the $H_f$ will naturally be low valued, so deterministic systems with a single frequency, such as a sine wave, have an entropy value of 0. For the case of wide distribution of frequencies, like for the white or pink noises, the value of $H_f$ will be much higher. 
For instance white Gaussian noise have the maximal $H_f$ on a given scale.

\begin{figure}[htbp]
    \centering
    \includegraphics[width=1.\linewidth]{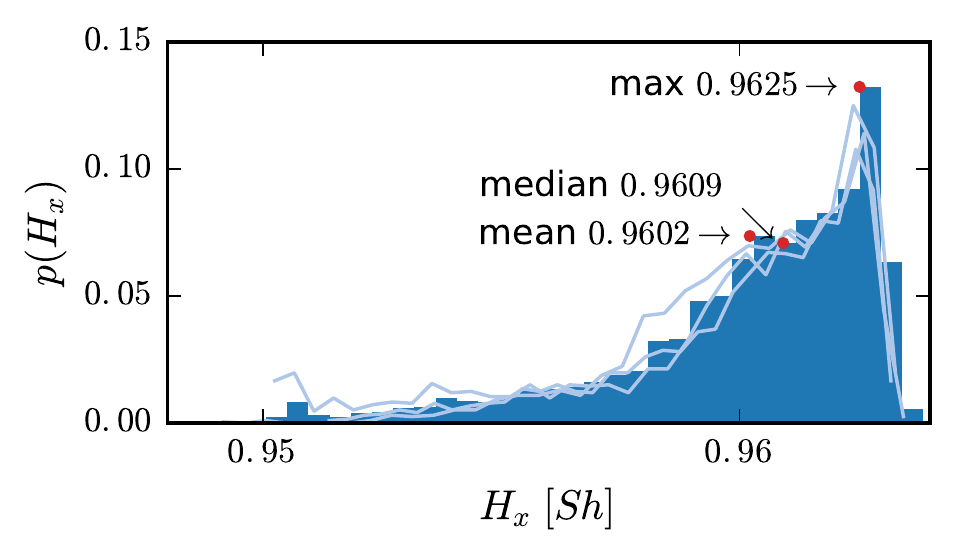}
    \caption{The distribution of Shannon entropy $p(H_x)$ calculated for all signals. Light solid lines represent histograms obtained separately for pre/post surgery recordings. Red dots mark the respective position of (left-to-right) mean, median and most probable value (max) of $H_x$.}
    \label{shanon}
\end{figure}

\section{Results}
\textbf{Shannon entropy.} We start with the estimation of the probability distribution $p(x)$ of the amplitudes of the signal. The simplest way to do so is to use the values of the signal itself and calculate their frequency. It is done by creating histograms of the amplitudes and normalise them by total number of samples $N$. Exemplary plots of the histograms can be seen in Fig. \ref{hist} for all $16$ channels at $5cm$ depth. 
At first glance all the characteristics look normal, although about 85\% of the analysed signals fail Lilliefors test \cite{lilliefors1967kolmogorov} and more than 90\% of signals fail Anderson-Darling normality test \cite{anderson1952asymptotic} at the significance level $\alpha=0.05$.

Next, for the estimated probabilities we calculate the Shannon entropy. In Fig. \ref{shanon} we depict the distribution of the obtained values. They range from 0.9492Sh to  0.9639Sh. The most common value is 0.9625Sh, while the average and median read 0.9602Sh and 0.9609Sh, respectively. 
All values are accurate to four significant digits. 
This characteristics stay similar for the pre/post stages of surgical treatment, c.f. light blue lines in Fig. \ref{shanon}.

\begin{figure}[htbp]
    \centering
    \includegraphics[width=\linewidth]{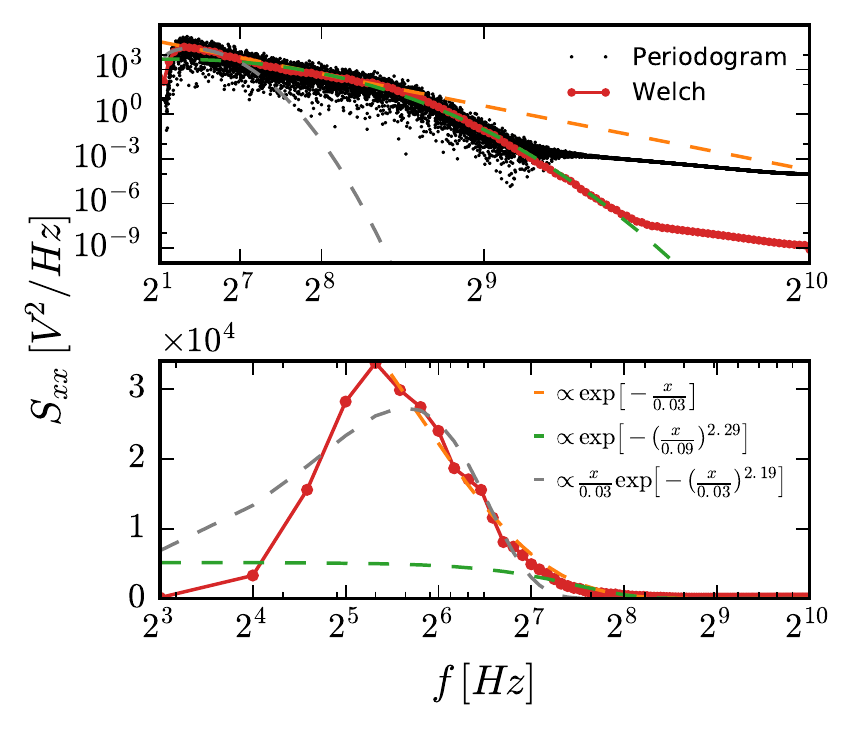}
    \caption{Exemplary power spectrum calculated for the raw sEMG data. Black dots denote periodogram. Red dots mark power spectrum calculated with the Welch's method. $S_{xx}$ for the slow dynamics can be approximated by the stretched exponential distribution of the Weibull form (dashed grey) or simple exponential (dashed orange) distribution. Intermediate and fast dynamics is represented by the stretched exponential distribution (dashed green), see text for details. 
    }
    \label{periodograms}
\end{figure}
\textbf{Spectral entropy.}
Similarly we start with the calculation of power spectra. For sEMG signals we found them rather complex where certain timescales can be described with different distribution functions, see Fig. \ref{periodograms} for details. 
Slow dynamics ($f < 2^7$, dashed grey) can be represented by the stretched exponential distribution of the Weibull form $\propto (x/\lambda_1) \exp[-(x/\lambda_1)^\beta_1]$, with the average values of $\lambda_1=0.0316 \pm 0.0095$ and $\beta_1=1.5708 \pm 0.4988$. 
On the other hand relatively slow dynamics ($50 < f < 200$, dashed orange) can also be represented by the simple exponential distribution $\propto \exp(- x / \lambda_2)$ with the average $\lambda_2=0.0325 \pm 0.0105$. 
$S_{xx}$ for the fast and intermediate frequencies ($f > 2^8$, dashed green) is given by the stretched exponential distribution $\propto \exp(- x / \lambda_3)^\beta_3$ with the average values of $\lambda_3=0.1109 \pm 0.0124$ and $\beta_3=2.6261 \pm 0.2037$.

\begin{figure}[htbp]
\centering
\includegraphics[scale=0.75]{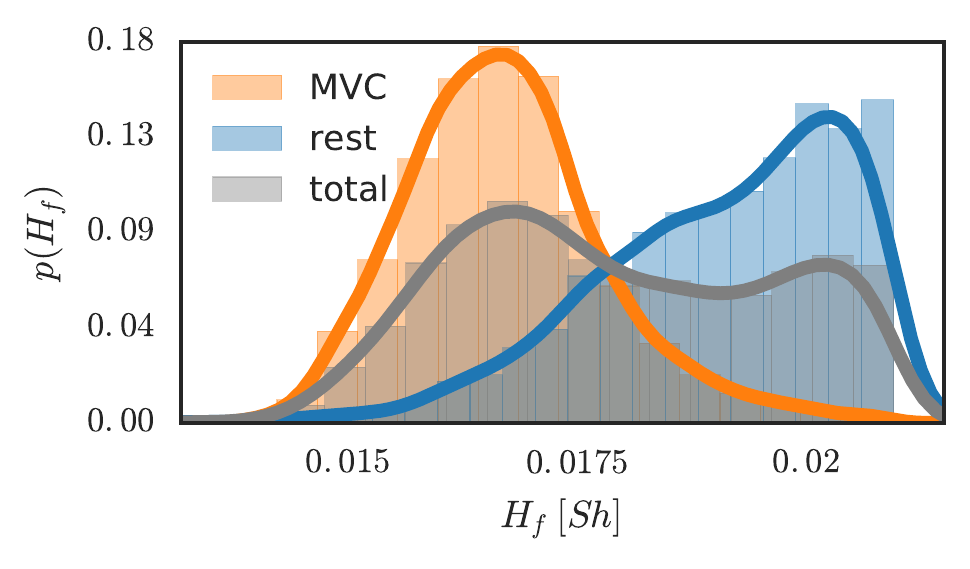}
\caption{The distributions of the spectral entropy $p(H_f)$ over the whole examined samples (gray) and separately for MVC (orange) and rest states (blue). Bar plots represent the histograms, while solid lines depict the kernel density estimation. Note shifted distribution of the MVC and rest states. See text for physiological explanation.}
\label{hf}
\end{figure}

In comparison with the probability based $H_x$, the typical values of $H_f$ are drastically lower, see Fig. \ref{hf}.
The average value of the spectral entropy for the total number of signals equals $\langle H_f \rangle = 0.0179 \pm 0.0017$. 
The distribution of the spectral entropy over the whole samples in question is bi--modal, c.f. grey histogram in Fig. \ref{hf}.
This can be explained once we consider separately the rest (blue) and MVC (orange) states. 
For the rest state the shape of the distribution of spectral entropy reminds the distribution of the Shannon entropy, with specific most common (maximum) state and the negative skewness, see Tab. \ref{setab}. 
The distribution of the $H_f$ for MVC is much closer to normal with the kurtosis 3.2969 and skewnes 0.2556.
In contrast to the $H_x$ its frequency domain equivalent also indicates a significant decrease of the average value $H_f$ for MVC. 
Both distribution are rather leptokurtic as both produce rather above average number of outliers than the standard normal distribution. 

This situation is partially justified by the concept of entropy as a measure of the diversity of the available states in the system. 
The reduction of these states is a consequence of the muscle contraction phenomenon itself. 
During the propagation of the action potential within the functional motor units the specific direction of the process is dominated, which automatically entails the decrease of the number of possible states that the system can choose.

\begin{table}[htbp]
\caption{The basic statistical description of the distributions of the spectral entropy in sEMG signals presented in Fig. \ref{hf} separately for rest and MVC states as well as for all the signals. All values are accurate to four significant digits. Note similar average values for all sets. Close-to-Gaussian distribution of $H_f$ for MVC state can be expressed with rather low skewness and similar kurtosis.}
\label{setab}
\centering
\begin{tabular}{lccc}
\toprule
{} &            total & rest & MVC \\
\midrule
mean  &     0.0179 & 0.0190 & 0.0167\\
std   &     0.0017 & 0.0015 & 0.0011 \\
skewness & -0.0894 & -1.4040 & 0.2556 \\
kurtosis & -0.2013 &  4.3347 & 3.2969 \\
\bottomrule
\end{tabular}
\end{table}

\section{Discussion and Conclusions}
The purpose of this work is to characterise the average information content of sEMG signals registered from external anal sphincter during multi--modal rectal cancer therapy. 
% We report the information for time and frequency... 
The data were examined in terms of application of the basic notions of information theory with a special effort devoted to the frequency distribution and the spectral entropy which is built upon it.
The spectral analysis itself brings a valuable information about the character of data. 
The power spectrum density for frequencies of the electrical activity of EAS is relatively wide, with the power that decrease exponentially for slow and even faster for the intermediate and higher frequency components.
For the slow dynamics there is clear frequency of around 50Hz for which $S_{xx}$ is maximal for all signals.

The standard probability based Shannon entropy did not provide any significant discrepancy for the signals measured before and after the surgery or for the relaxation vs MVC stages. 
This is partially a consequence of noisy character of sEMG data and similar probability distributions obtained for all the compared states. 
The broad power spectrum identified below the Nyquist limit should result in the relatively large values of $H_f$, which is not confirmed by our results. 
The effect of the drastically decreasing of $H_f$ and its very small values can be justified by the dominance of the narrow-banded slow dynamics and fast decreasing power spectrum for fast dynamics. 
Despite the a little amount of spectral information contained by the EAS activity, in contrast to the Shannon estimator, $H_f$ parameter clearly distinguishes the extreme states of muscle tension (relax vs MVC). 
{This contrast can also be found for multiscale $SampEn$ (MSE). For the relaxation state of muscle the value of MSE grows much faster to it's maximum value with the increase of the scale factor.} \cite{trybek2018sample}.
The anal sphincter muscle group resting activity contains a valuable amount of information which in frequency domain is greater than for the case of forced activity. 
This feature can be of importance for the clinicians in case of pre-surgery survey with sEMG. 

The main limitations of this study is due to the problem of inter–subject variability. 
The large diversity in distribution of EAS innervation zones, mainly caused by the high level of the individual asymmetries significantly affects the differences between the compared groups. 
That effect potentially has the greatest impact on very weak ability to distinguish the different states of treatment. 
Also the computational technique used has its internal limitations such as insensitivity to non--linear character of data.  
However, to our knowledge this is the first study which implements the information theory to the complex neuromuscular EAS activity and further investigations of the issue is still required. 
In our future research we intend to particularly concentrate on the methods that next to the statistical properties of data also taking into consideration the signal dynamics.

\end{document}